\documentclass[aoas,MSNbibl,nameyear,dvips]{arximspdf}
\usepackage{dcolumn}
\usepackage{graphicx}

\doi{10.1214/11-AOAS490}
\volume{5}
\issue{4}
\pubyear{2011}
\firstpage{2651}
\lastpage{2667}

\makeatletter

\setattribute{copyright}{owner}{\textup{In the Public Domain}}

\newcolumntype{d}[1]{D{.}{.}{#1}}

\makeatother

\begin{document}
\begin{frontmatter}

\title{A combined efficient design for biomarker data subject to a
limit of detection due to measuring instrument sensitivity\thanksref{T1}}
\runtitle{A combined efficient design subject to a limit of detection}

\thankstext{T1}{Supported by the Intramural Research Program of
the Epidemiology branch of the Eunice Kennedy Shriver National
Institute of Child Health and Human Development, NIH.}

\begin{aug}
\author[A]{\fnms{Enrique F.} \snm{Schisterman}\corref{}\ead[label=e1]{schistee@mail.nih.gov}},
\author[B]{\fnms{Albert} \snm{Vexler}\ead[label=e2]{avexler@buffalo.edu}},
\author[A]{\fnms{Aijun} \snm{Ye}\ead[label=e3]{yea2@mail.nih.gov}}\\
\and
\author[A]{\fnms{Neil J.} \snm{Perkins}\ead[label=e4]{perkinsn@mail.nih.gov}}
\runauthor{Schisterman, Vexler, Ye and Perkins}
\affiliation{Eunice Kennedy Shriver National Institute of Child
Health and Human Development, University of New York at Buffalo,
Eunice Kennedy Shriver National Institute of Child
Health and Human Development and
Eunice Kennedy Shriver National Institute of Child
Health and Human Development}
\address[A]{E. F. Schisterman\\
A. Ye\\
N. J. Perkins\\
Eunice Kennedy Shriver National Institute\\
\quad of Child Health and Human Development\\
6100 Executive Boulevard\\
Rockville, Maryland 20852\\
USA\\
\printead{e1}\\
\hphantom{E-mail: }\printead*{e3}\\
\hphantom{E-mail: }\printead*{e4}}
\address[B]{A. Vexler\\
University of New York at Buffalo\\
246 Farber Hall\\
3435 Main Street\\
Buffalo, New York 14214\\
USA\\
\printead{e2}} %adresu isvedimo komanda gale!
\end{aug}

% HISTORY:
\received{\smonth{12} \syear{2010}}
\revised{\smonth{6} \syear{2011}}

% ABSTRACT
%
\begin{abstract}
Pooling specimens, a well-accepted sampling strategy in biomedical
research, can be applied to reduce the cost of studying biomarkers.
Even if the cost of a single assay is not a major restriction in
evaluating biomarkers, pooling can be a powerful design that increases
the efficiency of estimation based on data that is censored due to an
instrument's lower limit of detection (LLOD). However, there are
situations when the pooling design strongly aggravates the detection
limit problem. To combine the benefits of pooled assays and individual
assays, hybrid designs that involve taking a sample of both pooled and
individual specimens have been proposed. We examine the efficiency of
these hybrid designs in estimating parameters of two systems subject to
a LLOD: (1) normally distributed biomarker with normally distributed
measurement error and pooling error; (2) Gamma distributed biomarker
with double exponentially distributed measurement error and pooling
error. Three-assay design and two-assay design with replicates are
applied to estimate the measurement and pooling error. The Maximum
likelihood method is used to estimate the parameters. We found that the
simple one-pool design, where all assays but one are random individuals
and a single pooled assay includes the remaining specimens, under
plausible conditions, is very efficient and can be recommended for
practical use.
\end{abstract}

% KEYWORDS
%

\end{frontmatter}

%s1 ###
\section{Introduction}

Epidemiological studies frequently investigate the relationship
between biomarkers and disease. In such studies, assaying
specimens for biomarkers can be expensive. For example, a single
assay to measure polychlorinated biphenyl (PCB)
costs between \$500 and \$1,000
[\citet{LWWSSLCGKPCB}]. %$^{3}$
The high cost severely constrains the number of assays that can be
performed in a study, thereby limiting the study's ability to
characterize a~biomarker-disease association.

Two study designs, the pooling design and the simple random sampling
design, have been proposed to reduce total assaying cost. Pooling
involves assaying only pooled, that is, physically mixed,
specimens [\citet{SBCOODNAPool}]. Each pooled specimen is
obtained by mixing pooling group size $p$ individual specimens
together, and each pooled specimen is assumed to contain an
amount of biomarker that is the mean of the amounts contained in its
constituent individual specimens [\citet{VLSLOD},
\citet{FRSLOD}, Schisterman et~al.
(\citeyear{SFRTAUC,SPLBYouden}), \citet{VLESLOD}]. %$^{1,5,11-13}$
Simple random sampling involves assaying only a simple random sample
of individual specimens [\citet{DDefect},
\citet{LSPool}, \citet{LSTPowerPool},
\citet{VSLROC}, \citet{WUPool}, \citet{ZGPool}]. %$^{1-10}$

Not only does cost hinder the characterization of a
biomarker-disease association, instrument sensitivity does as
well. An instrument may be unable to detect an amount of biomarker
below a certain level, the lower limit of detection (LLOD)
[\citet{VLSLOD}, \citet{MSVLPoolLOD},
\citet{VLESLOD}, \citet{SVWLLOD}].
%$^{1,2,13,14}$
Biomarker values above the LLOD are numerically determined, but
values below the LLOD are censored. Because instrument sensitivity is
an important issue in many areas such as occupational medicine and
epidemiology, LLOD issues have been extensively dealt with in the
biostatistical literature [\citet{SVWLLOD},
\citet{RCLOD}].
%$^{14,15}$

Investigations of the efficiencies of pooling and simple random
sampling in parameter estimation when data are subject to a LLOD
have been performed. \citet{MSVLPoolLOD} and \citet{VLSLOD} showed
that, in the context of biomarker mean and variance estimation,
there is always an interval of LLOD values for which pooling is more
efficient than simple random sampling and sometimes even more
efficient than assaying each and every individual specimen. This
phenomenon can be explained by the fact that, when a LLOD is below
the mean of a biomarker distribution, a pooled assay has a greater
chance of being above the LLOD than an individual assay
[\citet{SVWhenPool}]. \citet{MSVLPoolLOD} also showed that pooling
is more efficient than simple random sampling at estimating the area
under the receiver operating characteristic
curve (AUC) when the LLOD affects less than 50\% of the data. %$^{1}$
However, when the LLOD\vadjust{\goodbreak} is substantially greater than the mean of the
biomarker distribution, the pooling design is less efficient than
simple random sampling at estimating the AUC. Furthermore, the
reconstruction of individual assays' characteristics from pooled
data is generally a complex issue [\citet{VSLROC}].

The merits of pooling and simple random sampling led to the
consideration of hybrid designs, which combine pooling and simple
random sampling. Some randomly sampled individual specimens are
each assayed, and the remaining assays are pooled assays. The
efficiency of hybrid designs at parameter estimation has been
considered when data are not affected by a LLOD [\citet{SVMPHybrid}].
The present article extends previous work by examining the
efficiency of a variety of hybrid designs at estimating biomarker
distribution parameters and any assaying errors, when assays are
affected by a LLOD. When LLOD is present, ignoring missing or replacing
missing with a value might lead to severe bias. So it is important to
extend our previous work by including LLOD. Furthermore, we demonstrate
some hybrid designs under different situations in this article. We
consider the efficiency of hybrid designs
under various combinations of pooling error and measurement error.
Particularly, we are interested in a special case of the general hybrid
design, which we call the one-pool design, where all assays but one are
random individuals and a single pooled assay includes the remaining
specimens. This one-pool design is easy to execute in practice. Our
approaches can apply to the upper limit detection (ULOD) as well.

In the following sections we examine the efficiencies of hybrid
designs when data are subject to various errors and LLOD. Three-assay
design and two-assay design with replicates are applied to account for
the pooling error and measurement error. Three-assay design combines
one individual sampling group and two pooling groups with different
pooling size; while the two-assay design with replicates combines an
individual sampling group and one pooling group where each group is
measured in replicate. Both designs can be used to estimate the
parameters of the biomarker, measurement error and pooling error. The
variances of parameters are evaluated for both normally and Gamma
distributed biomarker levels. Last, we apply hybrid design to two
cases: (1) normally distributed data on cholesterol, a~coronary heart
disease biomarker and (2) Gamma distributed data on a chemokine
biomarker with double exponentially distributed measurement error and
pooling error.

%s2 ###
\section{Pooled-unpooled hybrid design subject to a LLOD}

In this section we describe a hybrid design, which combines assays on
individual specimens and assays on pooled specimens, when
assays are subject to a LLOD. Suppose we have $N$ uncorrelated
specimens $\{X_s, s=1,\ldots,N\}$, and we can perform only~$n$ assays. Let
$\alpha$ be the proportion of~$n$ that are assays of individual
specimens randomly sampled
from all individual specimens. When $\alpha=1$, only $n$ of the~$N$
specimens are used for a simple random sampling design. We measure
$\alpha n$ individual specimens $\{X_s, s=1,\ldots,\alpha n\}$ and
use the remaining $N - \alpha n$ individual specimens $\{X_s,s=\alpha
n+1,\ldots,N\}$ to create
$(1 - \alpha)n$ pooled specimens $\{X_i^{(p)},i=1,\ldots,(1-\alpha)n\}$.
Here we use subscript~$i$ to indicate assays. Ideally we would obtain
pooled measurements
\[
X_{i}^{(p)}=\frac{1}{p}\sum_{s=(i-1)p+\alpha n+1}^{ip+\alpha
n}X_{s},
\]
where $p$ is pooling group size, $p=[\frac{N-\alpha n}{(1-\alpha
)n}]$. Here $[x]$ is the integer round of a~quantity $x$. When $\alpha n=n-1$, we have one-pool design with $n-1$
individual assays $\{X_i, i=1,\ldots,n-1\}$ and 1 pooled assay $\{
X_1^{(p)}\}$. $\alpha_{\mathrm{one}\mbox{-}\mathrm{pool}}=1-\frac{1}{n}$ is the maximum of
$\alpha$ under hybrid design.

In this article we study the hybrid design in a realistic scenario
where assays have measurement error and pooling error as well as
subject to a~LLOD. A simple two-assay hybrid design composed of an
individual assay group and a pooled assay group is not enough to
estimate both measurement error and pooling error. We can apply two
approaches to estimate both errors: (1)~three-assay hybrid design and
(2) two-assay hybrid design with replicates.

%s2.1 ###
\subsection{Three-assay hybrid design}

A three-assay hybrid design consists of three different groups, an
individual group $Z^{(1)}$, a pooled group $Z^{(p_1)}$ of
pooling group size $p_1$, and a pooled group $Z^{(p_2)}$ of pooling
group size $p_2$. Let $\alpha$ be the fraction of assays
that are individual assays, and $\beta$ the
fraction of assays that are second pooled assays with pooling size
$p_2$. The numbers of the assays in each group are $n_1=\alpha n$,
$n_{p_1}=(1-\alpha-\beta) n$, and $n_{p_2}=
\beta n$, respectively. The total number of the specimens are $N=\alpha
n+(1-\alpha-\beta) np_{1}+\beta np_{2}$. Given $\beta$ and $p_2$, we
can obtain $p_1=[\frac{N-\alpha n-\beta n p_2}{(1-\alpha-\beta)
n}]$.
Due to the LLOD, each observation takes the following forms:
\[
Z_{i}^{(w)}= \cases{
X_{i}^{(w)}+\gamma
(w)e_{i}^{(p)}+e_{i}^{(m)}, &\quad
$X_{i}^{(w)}+\gamma
(w)e_{i}^{(p)}+e_{i}^{(m)}\geq \mbox{LLOD}$,
\vspace*{2pt}\cr
N/A, &\quad
$X_{i}^{(w)}+\gamma
(w)e_{i}^{p}+e_{i}^{(m)} < \mbox{LLOD}$,}
\]
where $w=1,p_1,p_2$ ($p_{1}\neq p_{2}$, since the three-assay design
reduces to the two-assay design when $p_{1}=p_{2}$), $i=1,\ldots,n_{w}$,
$X_i^{(1)}$ are the individual specimens, $e_{i}^{(m)}$ is measurement
error, $e_{i}^{(p)}$ is pooling error, and $\gamma(w)$ is a known
function such that $\gamma(1) = 0$. For simplicity, we assume $\gamma
(p_1)=\gamma(p_2)=1$. When $\beta=0$, three-assay design reduces to
two-assay design. When $\alpha n=n-1-\beta n$, we have one-pool design
with $n-1-\beta n$ individual assays $\{X_i, i=1,\ldots,n-1-\beta n\}$, 1
pooled assay $\{X_1^{(p_1)}\}$ with pooling size $p_1$, and $\beta n$
pooled assays $\{X_i^{(p_2)}, i=1,\ldots,\beta n\}$ with pooling size
$p_2$. We have $\alpha_{\mathrm{one}\mbox{-}\mathrm{pool}}=1-\beta-\frac{1}{n}$.
When $\beta=0$, three-assay design reduces to two-assay design, that
is, $\alpha_{\mathrm{one}\mbox{-}\mathrm{pool}}=1-\frac{1}{n}$.

%s2.2 ###
\subsection{Two-assay design with replicates}

Another approach to estimate pooling and measurement errors is
two-assay design with replicates. In practice, laboratories often
measure the assays twice. When a specimen is measured twice, for
individual samples, we have
\[
Z_{i1}^{(1)}=X_i+e_{i1}^{(m)}, \qquad  Z_{i2}^{(1)}=X_i+e_{i2}^{(m)},
\]
where $Z_{i1}^{(1)}$ and $Z_{i2}^{(1)}$ are measured values, $X$ is the
true value, and $e_{i1}^{(m)}$ and~$e_{i2}^{(m)}$ are measurement
errors. In practice, laboratories often use the average of $Z_{i1}^{1}$
and $Z_{i2}^{1}$ as the true biomarker value. We also have
%
%e1 ###
\begin{equation}\label{measureerror}
\Delta Z_i^{(1)}=Z_{i1}^{(1)}-Z_{i2}^{(1)}=e_1^{(m)}-e_{i2}^{(m)}.
\end{equation}
By fitting the distribution of $\Delta Z_i^{(1)}$, we can obtain the
parameter for measurement error $e^{(m)}$.
For pooled assays, we have
\[
Z_{i1}^{(p)}=X_i+e_1^{(m)}+e_1^{(p)},\qquad
Z_{i2}^{(p)}=X_i+e_{i2}^{(m)}+e_{i2}^{(p)},
\]
where $e_1^{(p)}$ and $e_{i2}^{(p)}$ are pooling errors. We also have
%
%e2 ###
\begin{equation}\label{poolingerror}
\Delta
Z^{(p)}=Z_{i1}^{(p)}-Z_{i2}^{(p)}=\bigl(e_1^{(m)}+e_1^{(p)}\bigr)-\bigl(e_{i2}^{(m)}+e_1^{(p)}\bigr).
\end{equation}
By fitting the distribution of $\Delta Z_i^{(p)}$, we can obtain the
parameter for the sum of measurement error and pooling error
$e^{(m)}+e^{(p)}$. After we obtain the estimates of the pooling and
measurement errors, we can use a two-assay design involving one
individual sampling group and only one pooling group to estimate the
parameters of the biomarker.

%s2.3 ###
\subsection{Maximum likelihood estimate}

The literature on limit of detection is largely maximum likelihood (ML)
due to a need to assume a distribution for the data that are
unmeasurable below the limit of detection. For insight below the limit
of detection, the distribution above is assessed and assumed consistent
below. ML estimation follows naturally after this. One simple way to
address the LLOD is to substitute a replacement value for unobservable
data. However, it will lead to biased assessment and it has been shown
that the best value is often $E[X|X<d]$ and required the same
assumption on the distribution below the limit of detection. In this
article, we use the ML method to handle LLOD data because it yields
asymptotically unbiased estimates of the parameters
[\citet{gupta1952}, \citet{chapman1956}]. We consider two cases: (1) normally
distributed biomarker with normally distributed measurement error and
pooling error, and (2) Gamma distributed biomarker with double
exponentially distributed measurement error and pooling error.

%s2.3.1 ###
\subsubsection{Normal distributed biomarker and errors}

Let the individual bio\-marker values be independently and identically
distributed as follows:
\[
X_{i}\sim N(\mu_{x},\sigma_{x}^{2}), \qquad i=1,\ldots,\alpha
n,\vadjust{\goodbreak}
\]
where $\alpha\in(0,1)$. By applying the pooling design based on
$N-\alpha n$ assays, ideally we would obtain pooled measurements
following normal distribution
\[
X_{i}^{(p)} \sim N
\biggl(\mu_{x},\frac{\sigma_{x}^{2}}{p}\biggr),\qquad
i=1,\ldots,(1-\alpha)n.
\]
We assume that the measurement error $e^{(m)}$ and pooling error
$e^{(p)}$ also follow independent normal distribution
\[
e_{i}^{(m)} \sim N(0, \sigma_m^2),\qquad
e_{i}^{(p)} \sim N(0, \sigma_p^2), \qquad  i=1,\ldots,n.
\]
The detailed likelihood function is available in the supplementary
material [Schisterman et~al. (\citeyear{Schistermanetal11}), Section~1].

%s2.3.2 ###
\subsubsection{Gamma distributed biomarker and double exponentially
distributed errors}

In certain situations, the distribution of the biomarker values is
skewed, and the normality assumptions cannot be applied. In these
circumstances, the Gamma distribution is a reasonable alternative.
Furthermore, the distribution of measurement and pooling errors can
vary by shape, and the normality assumptions are not always reasonable.
In these cases, double exponential distribution might be appropriate,
because it is symmetric and mean zero.
Suppose that the individual biomarker $X_i$ follows a Gamma distribution
\[
X_i \sim g(x;a,b)=\frac{1}{b^a \Gamma(a)}e^{-{x/b}}x^{a-1},\qquad
i=1,\ldots,\alpha n.
\]
For pooled assays with pooling size $p$, using the additive property of
the Gamma distribution, we have
\[
X_{i}^{(p)} \sim g(x; ap, b/p),  \qquad i=1,\ldots,(1-\alpha)n,
\]
and the measurement error and pooling error follow a double exponential
distribution with scale parameters $c$ and $d$, respectively,
\[
e_i^{(m)} \sim h(x;c)=\frac{1}{2c}e^{-{|x|/c}},\qquad
e_i^{(p)} \sim h(x;d)=\frac{1}{2d}e^{-{|x|/d}}
,  \qquad i=1,\ldots,n.
\]
The detailed likelihood function is available in the supplementary
material [Schisterman et~al. (\citeyear{Schistermanetal11}), Section 2].

%s2.4 ###
\subsection{Evaluation}

In this section we evaluate three cases: (1) normally distributed
biomarker with negligible measurement error and pooling error under
two-assay design, (2) normally distributed biomarker with normally
distributed measurement error and pooling error under three-assay
design, and (3) Gamma distributed biomarker and double exponentially
distributed measurement error and pooling error under two-assay design.

%s2.4.1 ###
\subsubsection{Normal case with negligible pooling error and
measurement error}\label{subnomp}

We are interested in the one-pool design, a special case of the hybrid
design, because it is simple and easily executed in practice. The
one-pool design fixes the $\alpha n=n-1$ individual sampling group,
leaving $(1-\alpha) n=1$ of the remaining $N-(n-1)$ specimens. We first
use a simple case with negligible pooling error and measurement error
to illustrate the efficiency of the one-pool design.

When\vspace*{1pt} random sampling and pooling are combined in the hybrid design,
the data consist of\vspace*{-1pt} individual and pooled observations
$\{Z_{1}^{(1)},\ldots,Z_{[\alpha
n]}^{(1)}$, $Z_{1}^{(p)},\ldots, Z_{[(1-\alpha)n]}^{(p)}\}$. If we
assume that the measurement error and pooling
error are negligible, that is, $e^{(m)}=0$ and $e^{(p)}=0$, the
three-assay design is reduced to a two-assay design ($\beta=0$). Each
observation takes the form
\[
Z_{i}^{(w)}=\cases{
X_{i}^{(w)}, &\quad $X_{i}^{(w)} \geq \mbox{LLOD}$, \vspace*{2pt}\cr
N/A, &\quad $X_{i}^{(w)} < \mbox{LLOD}$.}
\]
The log-likelihood function for normal distribution is a
function of only parameters $\mu_x$ and $\sigma_x$. To calculate the
MLEs of parameters
$\mu_{x}$ and $\sigma_{x}$, we solve the system of log-likelihood
first derivative\vspace*{2pt} equations \mbox{$\{ \frac{\partial\ell}{\partial
\mu_{x}}=0,\frac{\partial\ell}{\partial\sigma_{x}} =
0\}$}. Expressions for the log-likelihood equations and the
entries of Fisher information matrix I can be found in the
supplementary material [Schisterman et~al. (\citeyear{Schistermanetal11}), Section 3]. The
asymptotic variances of the estimators can be analyzed with respect
to $\alpha$ (the proportion of assays that are individual assays),
and an $\alpha$ that minimizes the variance of an MLE can be
proposed.

%
%f1 ###
\begin{figure}

\includegraphics{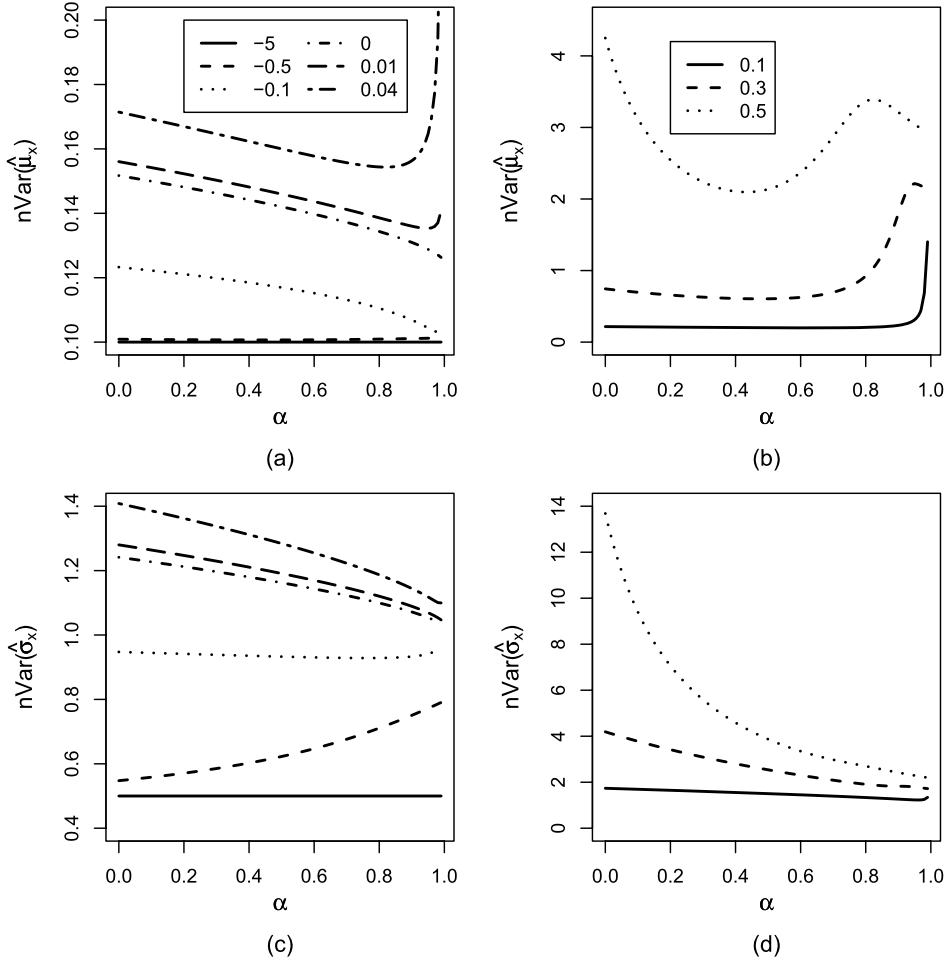}

\caption{$n\operatorname{Var}(\hat{\mu}_x)$ and
$n\operatorname{Var}(\hat{\sigma}_x)$ versus the proportion of individual assays to
the measured assays $\alpha$ in the absence
of measurement and pooling errors with $\mbox{LLOD}=-5, -0.5, -0.1$, 0, 0.01,
0.04, 0.1, 0.3 and 0.5 from bottom to top; $N=1\mbox{,}000$, $n=100$,
$\mu_x=0$
and $\sigma_x=1$.} \label{nomp}
\end{figure}

Figure \ref{nomp} illustrates the asymptotic variances $\operatorname{Var}(\hat{\mu
}_{x})$ and $\operatorname{Var}(\hat{\sigma}_{x})$ versus~$\alpha$ for $\mbox{LLOD}= -5, -0.5, -0.1, 0, 0.01, 0.04, 0.1, 0.3$
and 0.5
from bottom to top with $N = 1\mbox{,}000$, $n=100$, $\mu_{x} = 0$ and
$\sigma_{x} = 1$. Note that the rightmost point is at $\alpha
=(n-1)/n$, that is, one-pool design, rather than $\alpha=1$.

When LLOD is negligible (e.g., $\mbox{LLOD}=-5$), $\operatorname{Var}(\hat{\mu}_{x})$ is
approximately constant for $\alpha<
1$ in Figure \ref{nomp}(a). $\operatorname{Var}(\hat{\mu}_{x})$ increases with the
increase of LLOD. For $\mbox{LLOD}\leq
\mu_x$, $\operatorname{Var}(\mu_x)$ decreases as $\alpha$ increases, for example,
$\mbox{LLOD}=\mu_x$ (i.e., 0) and $\mu_x-0.1\sigma_x$ (i.e., $-0.1$).
$\operatorname{Var}(\mu_x)$ takes the minimum at $\alpha_{\mathrm{one}\mbox{-}\mathrm{pool}}=(n-1)/n=0.99$.
When $\mbox{LLOD} > \mu_x$, $\operatorname{Var}(\hat{\mu}_{x})$ takes a~minimum value at an
$0<\alpha<\alpha_{\mathrm{one}\mbox{-}\mathrm{pool}}$ as shown in Figure
\ref{nomp}(a) and
(b). A~hybrid design is more efficient than only measuring
pooled assays or only measuring individual assays. When $\mbox{LLOD}
<\mu_x$, the traditional pooling design ($\alpha=0$) is more
efficient than simple random sampling [\citet{VLSLOD},
\citet{MSVLPoolLOD}].
However, when a pooled-unpooled hybrid design is applicable, when
$\mbox{LLOD}\leq\mu_x$ and the objective is the estimate $\mu_x$, we
recommend a one-pool design given that pooling and measurement errors
are negligible. However, when $N$ is very large, pooling $N-(n-1)$
specimens might exceed the laboratory limitations.

Figure \ref{nomp}(c) shows $\operatorname{Var}(\hat{\sigma}_x)$ is approximately
constant as well when LLOD is absent (e.g., $\mbox{LLOD} =-5$). For
$\mbox{LLOD} <\mu
_x$ (e.g., $\mbox{LLOD} = \mu_x - 0.5\sigma_x$), pool design ($\alpha = 0$)
minimizes $\operatorname{Var}(\hat{\sigma}_x)$. For $\mbox{LLOD}\ge\mu_x$ (e.g., 0, 0.01,
0.04, 0.1, 0.3 and 0.5), $\operatorname{Var}(\hat{\sigma}_x)$ takes the minimum when
the one-pool design is used, as shown in Figure \ref{nomp}(c) and (d).

The traditional pooling design involves obtaining $n$ pooled assays
with pooling group size $p=N/n$. With this design, the variance of the
$\mu_{x}$-estimator
based on $n$ measurements of the pooled assays is $\sigma
_{x}^{2}/N$. For one-pool design with pooling group size $p=N-n+1$,
when the LLOD is not in effect, the MLE of $\mu_{x}$ based on the
combined data $%
\{Z_{1}^{(1)},\ldots,Z_{n-1}^{(1)},Z_{1}^{(N-n+1)}\}$ is the
following:
\begin{eqnarray*}
\hat{\mu}_{x}&=&\frac{1}{n-1+p}\Biggl(
\sum_{i=1}^{n-1}Z_{i}^{(1)}+pZ_{1}^{(p)}\Biggr) \\
&=&\frac{1}{N}\Biggl\{
\sum_{s=1}^{n-1}X_{s}+(N-n+1)\Biggl(\sum_{s=n}^{N}X_{s}/(N-n+1)
\Biggr)\Biggr\} .
\end{eqnarray*}
Thus, the one-pool design $\{Z_{1}^{(1)},\ldots,Z_{n-1}^{(1)}$,
$Z_{1}^{(N-n+1)}\}$ allows estimation of~$\mu_{x}$. $\operatorname{Var}(\hat{\mu}_x)$
is equivalent to that based on traditionally
pooled data $\{Z_{1}^{(N/n)},\ldots,\break Z_{n}^{(N/n)}\}$. This variance is
not equivalent to that based on a simple random sample of individual
assays $\{Z_{1}^{(1)},\ldots,Z_{n}^{(1)}\}$. The same conclusion can be
shown regarding the $\sigma^2_x$-estimation. This proposed one-pool
design is easier to execute than traditional
pooling. Moreover,\vspace*{1pt} if the parametric assumptions regarding the
sample distribution are rejected, the data $\{
Z_{1}^{(1)},\ldots,Z_{n-1}^{(1)}$, $Z_{1}^{(N-n+1)}\}$ can easily be
used to estimate the
unknown distribution, whereas reconstruction of the distribution
function of $X$ based on $\{Z_{1}^{(N/n)},\ldots,Z_{n}^{(N/n)}\}$ is a
very complicated problem [\citet{VSLROC}]. %$^{8}$
Even when the LLOD has a role, namely, when $\mbox{LLOD}\leq\mu_x$, as in
Figure \ref{nomp}(a), we can suggest the simple one-pool design.

%s2.4.2 ###
\subsubsection{Normal case with nonnegligible measurement error
and pooling error}

\begin{figure}

\includegraphics{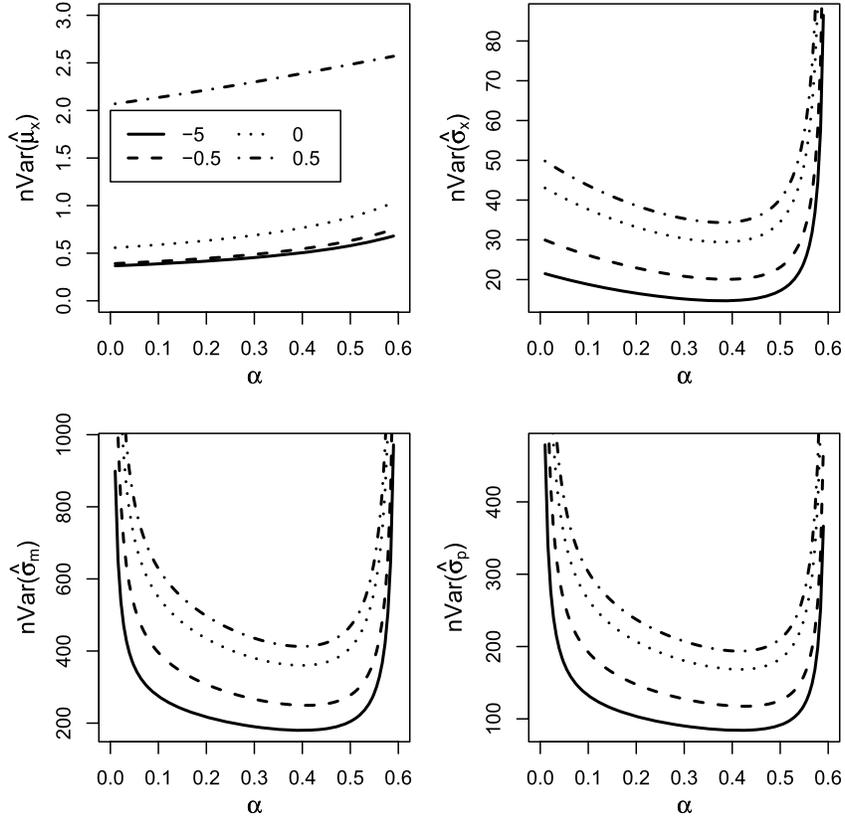}

\caption{$n\operatorname{Var}(\hat{\mu}_x)$, $n\operatorname{Var}(\hat{\sigma}_x)$,
$n\operatorname{Var}(\hat{\sigma}_m)$, and $n\operatorname{Var}(\hat{\sigma}_p)$ versus the
proportion of individual assays to the measured assays $\alpha$ in the
presence of
measurement and pooling errors under three-assay design with $\mbox{LLOD}=-5,
-0.5, 0$ and 0.5 from bottom to top; the proportion of the second
pooled assays $\beta=0.4$, pooling size $p_2=5$, $N=1\mbox{,}000$, $n=100$,
$\mu_x=0$, $\sigma_x=1$, $\sigma_m=0.3$ and $\sigma_p=0.4$.} \label{mp}
\end{figure}

When pooling error and measurement error are nonnegligible, one
approach to estimating the pooling and measurement errors is a
three-assay design, as mentioned at the beginning of this section. The
expressions for the normally distributed log-likelihood equations and
the entries of Fisher information matrix I can be found in the
supplementary material [Schisterman et~al.
(\citeyear{Schistermanetal11}), Section 4]. Figure \ref{mp} depicts the
%
%f2 ###
evolutions of $n\operatorname{Var}(\hat{\mu}_{x})$,
$n\operatorname{Var}(\hat{\sigma}_{x})$,
$n\operatorname{Var}(\hat{\sigma}_{p})$ and
$n\operatorname{Var}(\hat{\sigma}_m)$ with $N=1\mbox{,}000$, $n=100$,
$\sigma_{x} =1$, $\sigma _{p} = 0.3$ and $\sigma_m = 0.4$. The curves
from bottom to top are for $\mbox{LLOD}=-5, -0.5$, 0 and 0.5,
respectively. Because our hybrid design involves two pooling groups, we
set the proportion of the second pooling group $\beta=0.4$ and pooling
size $p_2=5$. Note that the rightmost point $\alpha=[(1-\beta)
n-1]/n=0.59$ is corresponding to the one-pool design that consists of
$(1-\beta) n-1=59$ individual assays, 1 pooled assay with pooling size
$p_1=741$, and $\beta n=40$ pooled assays with pooling size $p_2=5$.

As LLOD increases, $\operatorname{Var}(\hat{\mu}_{x})$, $\operatorname{Var}(\hat{\sigma}_{x})$,
$\operatorname{Var}(\hat{\sigma}_{m})$ and $\operatorname{Var}(\hat{\sigma}_{p})$ increase. $\operatorname{Var}(\hat
{\mu}_{x})$ increases as $\alpha$ increases, that is, the pooled design
minimizes $\operatorname{Var}(\hat{\mu}_{x})$. $\operatorname{Var}(\hat{\sigma}_{x})$, $\operatorname{Var}(\hat
{\sigma}_{m})$ and $\operatorname{Var}(\hat{\sigma}_{p})$ obtain the minimum under the
hybrid design. We provide R code as the supplementary material
[Schisterman et~al. (\citeyear{Schistermanetal11})] to calculate
$\operatorname{Var}(\hat{\mu}_{x})$, $\operatorname{Var}(\hat{\sigma}_{x})$, $\operatorname{Var}(\hat{\sigma}_{m})$
and $\operatorname{Var}(\hat{\sigma}_{p})$.

%s2.4.3 ###
\subsubsection{Gamma case with measurement error and pooling error}

In this subsection we study the situation with Gamma distributed
biomarker, double exponentially distributed measurement error and
pooling error by Monte Carlo simulation. Two-assay design can be used
when we know the variances of measurement error and pooling error. The
parameters for the Gamma distributed biomarker are $a=1.5$ and $b=0.1$.
So the mean of the individual biomarker is $E(X)=ab=0.15$, and the
variance $\operatorname{Var}(X)=ab^2=0.015$. The parameters for double exponentially
distributed measurement error and pooling error are $c=0.02$ and
$d=0.03$, respectively. Both errors are mean zero and the variance of
measurement error is $\operatorname{Var}(e^{(m)})=2c^2=0.0008$, and the variance of
pooling error $\operatorname{Var}(e^{(p)})=2d^2=0.00018$. The number of specimens is
$N= 1\mbox{,}000$ and the number of assays is $n= 100$. 1,000 simulations were
performed to evaluate $\operatorname{Var}(\hat{a})$ and $\operatorname{Var}(\hat{b})$ at $\alpha =
0$, 0.2, 0.4, 0.6, 0.8 and 0.99, subject to $\mbox{LLOD}=0.02, 0.05, 0.1$
and~0.15.\vadjust{\goodbreak}

\begin{figure}

\includegraphics{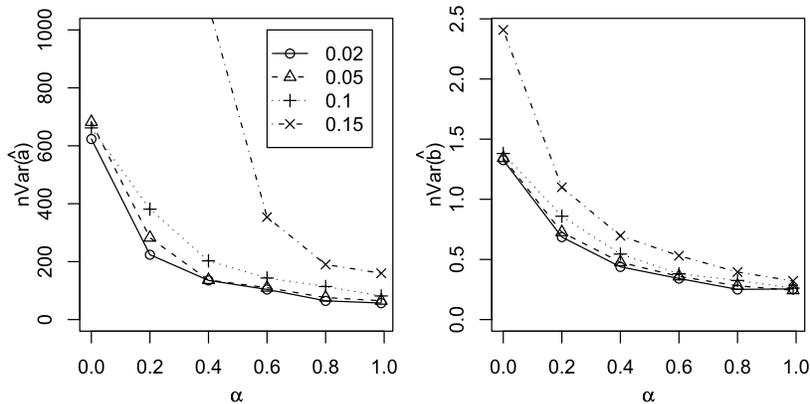}

\caption{$n\operatorname{Var}(\hat{a})$ and $n\operatorname{Var}(\hat{b})$ versus the proportion of
individual assays to the measured assays $\alpha$ for
simulated Gamma distributed biomarkers with double exponentially
distributed measurement and pooling errors with $N=1\mbox{,}000$, $n=100$, $a=1.5$,
$b=0.1$, $c=0.02$, $d=0.03$, $\mbox{LLOD}=0.02, 0.05, 0.1$ and 0.15.} \label{simu}
\vspace*{-3pt}
\end{figure}

The simulation results are presented in Figure \ref{simu}.
$\operatorname{Var}(\hat{a})$ and $\operatorname{Var}(\hat{b})$
increases with the increase of LLOD. Both $\operatorname{Var}(\hat{a})$
and $\operatorname{Var}(\hat{b})$ decrease with the increase of~$\alpha$.
%
%f3 ###
They are minimized under the one-pool design ($\alpha=0.99$). When
$\mbox{LLOD} < E(X)$, $\operatorname{Var}(\hat{a})$ does not change much with the increase of
$\alpha$. However, when $\mbox{LLOD} = E(X)$, $\operatorname{Var}(\hat{a})$ becomes
significantly larger, especially when $\alpha$ is small. It is
five-fold larger than with other LLOD values for pool design ($\alpha
=0$). $\operatorname{Bias}(\hat{a})$ and $\operatorname{Bias}(\hat{b})$ for finite sample size are
presented in Section 5 of the supplementary material [Schisterman et~al. (\citeyear{Schistermanetal11})]. They are
relatively small except for large LLOD, for example, $\mbox{LLOD} = 0.15$
(61\%
missing for individual sampling).

%s3 ###
\section{Application}

%s3.1 ###
\vspace*{-3pt}\subsection{Normally distributed biomarker with negligible
measurement and pooling errors}\vspace*{-3pt}

In order to investigate the efficiency of the hybrid design, we
bootstrapped by using real data from a study of biomarkers of coronary
heart disease. In this study, cholesterol level, a biomarker for
coronary heart disease, was measured for $40$ individuals that had a
normal rest electrocardiogram, were free of symptoms, and had no
previous cardiovascular procedures or myocardial infarctions. The
mean of the individual biomarker assays is $205.53$ mg/dl and the
standard deviation is $42.29$ mg/dl. The Shapiro--Wilk test for
normality suggests that the individual assays follow a normal
distribution.

We assume that we have $N = 40$ specimens, we can only afford to
perform $n = 20$ assays,\vadjust{\goodbreak} and the measurement error and pooling error
are negligible. Artificial $\mbox{LLOD}= 0, 150, 170, 180, 200, 205$ and 210
are applied to the cholesterol data. We evaluated six designs,
involving $\alpha$ values from Table \ref{tablenormal}. The rightmost
one ($\alpha=0.95$) is a one-pool design. To generate the pooled data
with different pooling size $p$, we pooled the individual assays
together, and used the average values as the measured values of the
pooled assays. Then we combined the unpooled
and simulated pooled data, and applied the methodology for two-assay design
with negligible measurement and pooling error case in Section
\ref{subnomp} to calculate the maximum likelihood estimate of $\mu
_{x}$. This procedure is repeated $100\mbox{,}000$ times to obtain
$\operatorname{Var}(\hat{\mu}_{x})$.

%
%t1 ###
\begin{table}
\caption{Parameters used for normally distributed biomarker ignoring
errors with number of samples $N = 40$ and the number of assays $n = 20$}
\label{tablenormal}
\vspace*{-3pt}
\begin{tabular*}{\tablewidth}{@{\extracolsep{\fill}}l c c c c c c @{}}
\hline
$\bolds{\alpha}$ & \textbf{0} & \textbf{0.5} & \textbf{0.75} &
\textbf{0.8} & \textbf{0.9} & \textbf{0.95} \\
\hline
Number of individual assays & \hphantom{0}0 & 10 & 15 & 16 & 18 & 19 \\
Number of pooled assays & 20 & 10 & \hphantom{0}5 & \hphantom{0}4 & \hphantom{0}2 & \hphantom{0}1 \\
Pooling size $p$ & \hphantom{0}2 & \hphantom{0}3 & \hphantom{0}5 & \hphantom{0}6 & 11 & 21 \\
\hline
\end{tabular*}
\vspace*{-3pt}
\end{table}

%
%f4 ###
\begin{figure}[b]
\vspace*{-3pt}
\includegraphics{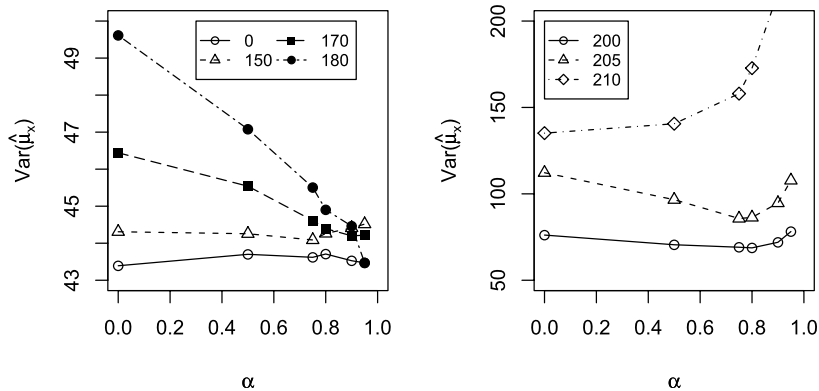}
\vspace*{-3pt}
\caption{$\operatorname{Var}(\hat{\mu}_x)$ versus the proportion of individual assays
to the measured assays $\alpha$ by
bootstrapping with $N=40$, $n=20$, $\mbox{LLOD}=0$, 150, 170, 180, 200, 205 and
210.} \label{bootstrap1}
\end{figure}

The results are shown in Figure
\ref{bootstrap1}. When $\mbox{LLOD}<\hat{\mu}_{x}-\hat{\sigma}_x$ (e.g., 0
and 150),
$\operatorname{Var}(\hat{\mu}_x)$ is approximately a constant. When $\hat{\mu}
_{x}-\hat{\sigma}_x< \mbox{LLOD} <\hat{\mu}_{x}$ (e.g., 170 and 180),
$\operatorname{Var}(\hat{\mu}_x)$ decreases as $\alpha$ increases. The minimum
is obtained under the one-pool design. When LLOD is close to $\hat{\mu}
_{x}$ (e.g., 200 and 205), $\operatorname{Var}(\hat{\mu}_x)$ takes the minimum at
$0<\alpha<1$. A hybrid design is favorable. Although the one-pool
design does not give the minimum, $\operatorname{Var}(\hat{\mu}_x)$ for the
one-pool design (78.2 for $\mbox{LLOD} = 200$) is close to the minimum (68.7).
Due to the simplicity of
design, one-pool design can be recommended. When $\mbox{LLOD} >\hat{\mu}_{x}$
(e.g., 210), $\operatorname{Var}(\hat{\mu}_x)$ increases as $\alpha$ increases. The
maximum of $\operatorname{Var}(\hat{\mu}_x)$ is obtained under one-pool design.\vadjust{\goodbreak}

%s3.2 ###
\subsection{Gamma distributed biomarker with double exponentially
distributed measurement error and pooling error}

In this subsection we exemplified the two-assay design with replicates
using real data from a study of chemokine biomarker monocyte
chemotactic protein-1 (MCP-1). MCP-1 plays a role in a variety of
pathological conditions such as inflammatory and immune reactions.
Assays are measured in different plates. Each plate has its own LLOD.
In this article we use only the data from the plates with $\mbox{LLOD} = 0.016$,
because our model requires the same LLOD. Each plate was measured
twice. There are 99 individual sampling assays, and 45 pooled assays
with $p=2$. The mean of the individual sampling assays is 0.189, and
the standard deviation is 0.183. The measurement errors can be
calculated by the difference of individual sampling assays [see
(\ref{measureerror})], and the pooling errors can be
calculated by the difference of pooled assays; see (\ref
{poolingerror}). We used the R package VGAM [\citet{vgam}] to fit the
difference of individual replicates $\Delta Z^{(1)}$ to obtain the
estimate of parameter $c$. Then we fit the difference of pooled
replicates $\Delta Z^{(p)}$, which follows a double exponential
distribution with parameter~$e$. The estimated variances of measurement
error and pooling error can be obtained by
\begin{eqnarray*}
\widehat{\operatorname{Var}}\bigl(e^{(m)}\bigr) &=& \frac{\widehat{\operatorname{Var}}(\Delta Z^{(1)})}{2},\\
\widehat{\operatorname{Var}}\bigl(e^{(p)}\bigr) &=& \frac{\widehat{\operatorname{Var}}(\Delta Z^{(p)})
-\widehat{\operatorname{Var}}(\Delta Z^{(1)})}{2}.
\end{eqnarray*}
After we obtained the estimates of the variances of pooling error and
measurement error, we used one individual sampling group and one
pooling group to estimate the other parameters, for example, $a$ and
$b$ of the Gamma distributed biomarker.

%
%f5 ###
\begin{figure}

\includegraphics{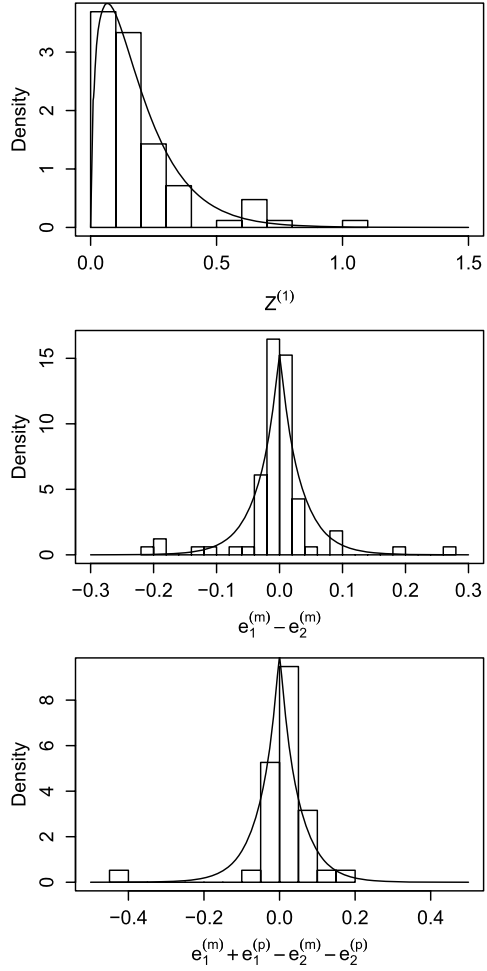}

\caption{Histograms of individual biomarker $Z^{(1)}$, difference of
measurement error $e_1^{(m)}-e_{i2}^{(m)}$ and difference of the sum of
measurement error and pooling error $(e_1^{(m)}+e_1^{(p)}
)-(e_{i2}^{(m)}+e_{i2}^{(p)})$.} \label{hist}
\end{figure}

The histograms of individual biomarker $Z^{(1)}$, difference of
measurement error $e_1^{(m)}-e_2^{(m)}$ and difference of the sum of
measurement error and pooling error
$(e_1^{(m)}+e_1^{(p)})-(e_2^{(m)}+e_2^{(p)})$ are illustrated in Figure
\ref{hist}. The fitting curves are generated by the parameters
estimated by the R package VGAM. The estimated parameters are presented
in Table \ref{tablepara}. For double exponential distribution, the
estimated variance is $2s^2$, where $s$ is the scale parameter of
double exponential distribution.
The estimated $\widehat{\operatorname{Var}}(e^{(m)})$ and $\widehat{\operatorname{Var}}(e^{(p)})$ are
presented in Table \ref{tablepara} as well as their corresponding scale
parameters. For Gamma distribution, the estimated mean is $ab$ and the
estimated variance is~$ab^2$. Table \ref{tablepara} shows that the
sample variances are very close to the estimated variances. The fitting
curves in Figure \ref{hist} fit the histogram quite well.\looseness=-1

For fixed $N$ and $n$, we need to vary the pooling size $p$ to vary
$\alpha$. However, we only have individual unpooled data and pooled
data with pooling size $p=2$. So we pool the $p=2$ pooled assays
together to generate the data with different pooling size. Because we
want to include the measurement error and\vadjust{\goodbreak} pooling error in the pooled
assays, we used pooled assays rather than individual sampling assays to
generate pooled assays with different pooling size. For example,
\begin{eqnarray*}
Z^{(p=4)}&=&\frac{1}{2}\bigl(Z_{i1}^{(p=2)}+Z_{i2}^{(p=2)}\bigr) \\
&=&\frac{1}{2}\biggl(\frac{X_1+X_2}{2}+e_1^{(p)}+e_1^{(m)}+
\frac{X_3+X_4}{2}+e_{i2}^{(p)}+e_{i2}^{(m)} \biggr) \\
&=&\frac{1}{4}(X_1+X_2+X_3+X_4)+\frac{1}{2}
\bigl(e_1^{(p)}+e_{i2}^{(p)}\bigr)+
\frac{1}{2}\bigl(e_1^{(m)}+e_{i2}^{(m)}\bigr).
\end{eqnarray*}
Then we combined individual unpooled data, and measured ($p=2$) or
simulated ($p>2$) pooled data to generate a hybrid design. The pooling
sizes we used are presented in Table \ref{tablegamma}. We assume that
we have $N = 79$ or 80 specimens, and can only afford to
perform $n = 40$ assays. Besides the true $\mbox{LLOD} = 0.016$, additional
$\mbox{LLOD}
= 0.05, 0.1$ and 0.15 are applied to evaluate the influence of LLOD.

%
%t2 ###
\begin{table}
\tabcolsep=0pt
\caption{The estimates of the parameters for individual biomarker
$Z^{(1)}$, difference of measurement error~$e_1^{(m)}-e_{i2}^{(m)}$,
difference of the sum of measurement error and pooling error
$(e_1^{(m)}+e_1^{(p)})-(e_{i2}^{(m)}+e_{i2}^{(p)})$, measurement error
$e^{(m)}$ and pooling error $e^{(p)}$. Here $a$ and $b$ are the shape
and scale parameters of the Gamma distribution, respectively, $s$
are the
scale parameters of double exponential distribution}
\label{tablepara}
\begin{tabular*}{\tablewidth}{@{\extracolsep{\fill}}l c c c d{1.3} d{2.4} d{1.4} d{1.4}@{}}
\hline
& & & & \multicolumn{2}{c}{\textbf{Mean}} & \multicolumn{2}{c@{}}{\textbf{Variance}} \\
[-4pt]
& & & & \multicolumn{2}{c}{\hrulefill} &
\multicolumn{2}{c@{}}{\hrulefill}\\
& $\bolds{a}$ & $\bolds{b}$ & $\bolds{s}$
& \multicolumn{1}{c}{\textbf{Estimated}} & \multicolumn{1}{c}{\textbf{Sample}} & \multicolumn{1}{c}{\textbf{Estimated}} &
\multicolumn{1}{c@{}}{\textbf{Sample}} \\
\hline
$Z^{(1)}$ & 1.54 & 0.12 & & 0.189 & 0.189 & 0.023 & 0.034 \\[2pt]
$e_1^{(m)}-e_{i2}^{(m)}$ & & & 0.033 & 0 & -0.0034 & 0.0022 & 0.0029 \\[2pt]
$e_1^{(m)}+e_1^{(p)}-e_{i2}^{(m)}-e_{i2}^{(p)}$ & & & 0.050 & 0 &
0.012 & 0.0051 & 0.0059 \\[2pt]
$e^{(m)}$ & & & 0.023 & & & 0.0011 & \\[2pt]
$e^{(p)}$ & & & 0.027 & & & 0.0015 & \\
\hline
\end{tabular*}
\end{table}

%
%t3 ###
\begin{table}[b]
\caption{Parameters used for the Gamma distributed biomarker with
double exponentially distributed errors and the number of assays $n = 40$}
\label{tablegamma}
\begin{tabular*}{\tablewidth}{@{\extracolsep{\fill}}l c c c c c@{}}
\hline
$\bolds{\alpha}$ & \textbf{0} & \textbf{0.675} & \textbf{0.8}
& \textbf{0.925} & \textbf{0.975} \\
\hline
Number of individual assays & \hphantom{0}0 & 27 & 32 & 37 & 39 \\
Number of pooled assays & 40 & 13 & \hphantom{0}8 & \hphantom{0}3 & \hphantom{0}1 \\
Pooling size $p$ & \hphantom{0}2 & \hphantom{0}4 & \hphantom{0}6 & 14 & 40 \\
Number of samples $N$ & 80 & 79 & 80 & 79 & 79 \\
\hline
\end{tabular*}
\end{table}

%
%f6 ###
\begin{figure}

\includegraphics{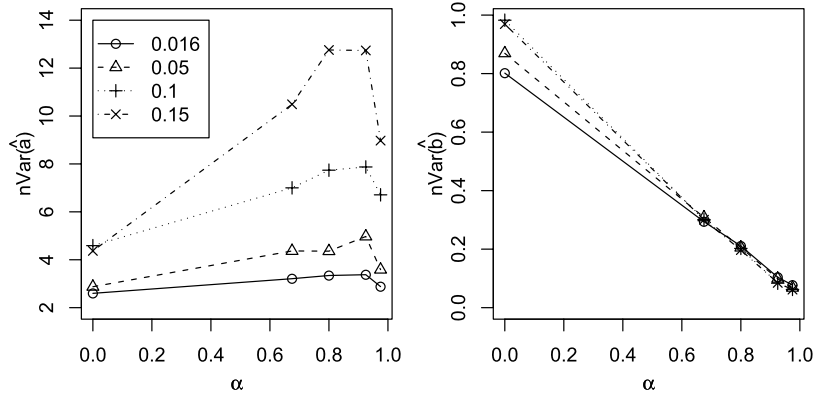}

\caption{$n\operatorname{Var}(\hat{a})$ and $n\operatorname{Var}(\hat{b})$ versus the proportion of
individual assays to the measured assays $\alpha$ by
bootstrapping with $N=79$ or 80, $n=40$, $\mbox{LOD}=0.016, 0.05, 0.1$ and 0.15.}
\label{bootstrap2}
\end{figure}

The results are illustrated in Figure \ref{bootstrap2}. As $\alpha$ increases,
$\operatorname{Var}(\hat{a})$ increases then decreases at the one-pool
design ($\alpha=0.975$). One-pool design gives the second minimum.
This tendency is different from the simulation result, where
$\operatorname{Var}(\hat{a})$ decreases as $\alpha$ increases, and the minimum is
reached under one-pool design. When LLOD is very small (i.e., $\mbox{LLOD} =
0.016$), $\operatorname{Var}(\hat{a})$ does not change much. $\operatorname{Var}(\hat{b})$
decreases as $\alpha$ increases, which is consistent with the
simulation result.

%s4 ###
\section{Summary and discussion}

Although the pooling design can increase the efficiency of
estimation from data subject to a LLOD, there are situations when the
pooling design strongly aggravates the detection limit\vadjust{\goodbreak} problem. A~hybrid design was proposed in order to gain benefits from both
individual assays and pooled assays
[\citet{SVMPHybrid}]. %$^{16}$

In this article we present methodology for determining a hybrid design
that most efficiently estimates parameters from data
subject to measurement error, pooling error and a limit of detection.
Efficiency is gauged by the variance of a maximum likelihood
estimator of a parameter. We demonstrated the asymptotic MLE variances
as functions of the proportion of individual assays to the measured
assays. To estimate both measurement error and pooling error, a
three-assay design or a two-assay design with replicates is needed. We
examined two cases: one is with the normally distributed biomarker and
errors, the other is with the Gamma distributed biomarker and double
exponentially distributed errors.

Under the condition that we have $N$ specimens and we can only perform
$n <
N$ assays, we evaluated the efficiency of the one-pool hybrid design,
which involves assaying $n-1$ individual specimens and one pooled
sample of the remaining $N-(n-1)$ individual specimens. When
measurement error and pooling error are negligible, for the normally
distributed biomarker, one-pool design minimizes $\operatorname{Var}(\hat{\mu}_x)$ for
$\mbox{LLOD} \le\mu_x$ and $\operatorname{Var}(\hat{\sigma}_x)$ for $\mbox{LLOD} > \mu_x$. When
measurement error and pooling error are in effect, the pooled design
minimizes $\operatorname{Var}(\hat{\mu}_{x})$, while the hybrid design minimize
$\operatorname{Var}(\hat{\sigma}_{x})$, $\operatorname{Var}(\hat{\sigma}_{m})$ and $\operatorname{Var}(\hat{\sigma
}_{p})$. The $\alpha$ value corresponding to the minimum can be
obtained by the~R code that we provided as the supplementary material [Schisterman et~al. (\citeyear{Schistermanetal11})]. Note
that, in practice, our interest is in $\mu_x$, $\sigma_x$, and not in
$\sigma_p$ or $\sigma_m$. The simulation result shows that it minimizes
both $\operatorname{Var}(\hat{a})$ and $\operatorname{Var}(\hat{b})$ for Gamma distribution under
complex measurement error and pooling error assumptions. Hence, under
the circumstances described above, when one seeks to avoid more
complicated procedures for determining and executing a potentially more
efficient hybrid design, the one-pool hybrid
design is an efficient and easily implemented alternative to a
simple random sample of individual assays.

\section*{Acknowledgments}

The authors thank the Editor, reviewers, Qian Zhang and Sonya Dasharathy
for their valuable comments, and Dr. Brian Whitcomb for the chemokines
data.

\begin{supplement}%[id=suppA]
\stitle{R code and detailed derivations}
\slink[doi]{10.1214/11-AOAS490SUPP} %[doi,text={...}] - jei reikia
%suskaldyti doi
\slink[url]{http://lib.stat.cmu.edu/aoas/490/supplement.pdf}
\sdatatype{.pdf}
\sdescription{R~code used to calculate
$n\operatorname{Var}(\hat{\mu}_x)$,
$n\operatorname{Var}(\hat{\sigma}_x)$,
$n\operatorname{Var}(\hat{\sigma}_m)$ and
$n\operatorname{Var}(\hat{\sigma}_p)$. Detailed derivation of maximum
likelihood estimates and the Fisher information matrix.}
\end{supplement}

% imsref loaded by lrinkeviciute, 2011-09-29 09:47:24
%
% imsref loaded by lrinkeviciute, 2011-09-29 09:51:51
% imsref loaded by lrinkeviciute, 2011-09-29 10:44:58
% imsref loaded by lrinkeviciute, 2011-09-29 11:11:32

\printaddresses

\end{document}